\documentclass{ifacconf}

\usepackage{graphicx}      % include this line if your document contains figures
\usepackage{natbib}        % required for bibliography

\usepackage{IEEEtrantools}

\usepackage{amsmath,amsfonts,bm}

\newcommand{\q}{\rangle}
\newcommand{\p}{\langle}
\begin{document}
\begin{frontmatter}

\title{Scaling up reservoir engineering for error-correcting codes\thanksref{footnoteinfo}} 

\thanks[footnoteinfo]{This work has been supported by Army Research Office (ARO) under Grant No. W911NF-18-1-0212, and by the ANR project HAMROQS.}

\author[First]{Vincent Martin} 
\author[First]{Alain Sarlette}

\address[First]{QUANTIC lab, INRIA, 2 Rue Simone IFF, 75012 Paris, France (e-mail: vincent.a.martin@inria.fr; alain.sarlette@inria.fr).}
%\address[Second]{2 Rue Simone IFF, 75012 Paris France (e-mail: alain.sarlette@inria.fr).}

\begin{abstract}                
Error-correcting codes are usually envisioned to counter errors by operating unitary corrections depending on the projective measurement results of some syndrome observables. We here propose a way to use them in a more integrated way, where the error correction is applied continuously and autonomously by an engineered environment. We focus on a proposal for the repetition code that counters bit-flip errors, and how to scale up the network encoding a logical quantum bit, towards stronger information protection. The challenge has been to devise a network architecture which allows to autonomously correct higher-order errors, while remaining realistic towards experimental realization by avoiding all-to-all or all-to-one coupling.
\end{abstract}


\end{frontmatter}
%===============================================================================

\section{Introduction}
Since the development of Shor's algorithm for factoring integer numbers (\cite{365700}), quantum systems are viewed as a promising tool to process information faster than classical computers. However, the uncontrolled effects of the environment on quantum systems cause so-called decoherence, degrading their quantum properties. Protecting information from such decoherence in order to build a memory for a quantum computer (see \cite{nielsen2010quantum} for an introductory textbook) is therefore a major challenge of quantum eningeering. A redundant encoding of information allows, by repeating comparative measurements, to estimate the errors that occur and thereby protect logical information. Such Quantum Error Correction (QEC, see also \cite{nielsen2010quantum}) has been proposed from the very beginning of the field.
The usual proposal to achieve it is to measure so-called syndrome observables projectively. To correct bit flips for instance, the syndromes are parities between each pair of neighboring qubits. A parity mismatch indicates an error, and the most likely initial state can be established on probability grounds.% A continuous-time version of this procedure has also been proposed (\cite{PhysRevA.65.042301}).

Reservoir engineering is a way to stabilize a system without measurement and feedback computation, by directly implementing  the feedback loop into the physical system (\cite{PhysRevLett.77.4728}, \cite{PhysRevLett.110.120501}). This is done by carefully crafting the Hamiltonian couplings between different subsystems, including dissipative (decoherence) channels, in order to constitute an effective engineered environment for the target system; the procedure is comparable to the classical Watt governor. The upside of this method is a great simplification of the experimental settings: the device works by itself at the quantum hardware level, without the need for online feedback with classical signals. This allows a much better physical isolation from possible external perturbations. The challenging part is of course to engineer such a physical feedback loop. Applying this method to QEC would be an interesting alternative for efficient protection of quantum information. The scope of the present paper is to study how such reservoir engineering for QEC can be scaled.

We start from the work of \cite{cohen:tel-01545186}, which aims to stabilize a logical qubit composed of 3 physical qubits in order to increase the memory fidelity under perturbations inducing physical bit flip errors. We first describe briefly the associated physical system and the resulting error correction protocol, before showing how it can be improved, scaling up the order of information protection, by coupling several of these systems into a network of 9 physical qubits. The new method and pattern proposed here is modular and can be used to further extend the code.

\section{The three qubit repetition code}

In this section, we recall the basic building block that will be used afterwards, as developed in \cite{cohen:tel-01545186}. Insights on the calculations can be found in \cite{Leghtas853}.

\subsection{Description of the system}

We will work on a system composed of 3 qubits each living in a two dimensional space (basis state $|0\q$ or $|1\q$) and of 3 harmonic oscillators (cavities) that can technically be considered as also only living on the two fist energy levels, as they will be strongly dissipative. Indeed, their role is to evacuate entropy from the system. The qubits in contrast contain the logical information, encoded on the two basis states $|000\q$ and $|111\q$ of their joint state space.
We will note $\mathbf{a_i}$ the annihilation operator on i-th cavity, $\boldsymbol{\sigma}^x_i$ and $\boldsymbol{\sigma}^z_i$ the Pauli operator on i-th qubit. Qubit decay and excitation correspond respectively to the operators $\boldsymbol{\sigma}^-_i = |0\rangle\langle 1 |_i$ and $\boldsymbol{\sigma}^+_i = |1\rangle \langle 0|_i$. The Lindblad superoperator expressing open quantum dynamics is defined by $\mathcal{D}_{\mathbf{X}}(\rho) = \mathbf{X} \rho \mathbf{X}^\dagger - \frac{\mathbf{X}^\dagger \mathbf{X} \rho + \rho \mathbf{X}^\dagger \mathbf{X}}{2}$, and h.c. stands for hermitian conjugate of the preceding terms.

Our system is described by the Lindblad equation:
\begin{equation}\label{eq:Lindbladian}
\frac{d \rho}{d t} = -i[\mathbf{H}(t), \rho] + \sum_{i=1}^3 \kappa \mathcal{D}_{\mathbf{a_i}}(\rho) + \sum_{i=1}^3 \gamma \mathcal{D}_{\boldsymbol{\sigma}^x_i}(\rho) 
\end{equation}
where $\rho$ is the state of the whole system (3 qubits and 3 cavities), $\kappa \mathcal{L}_{\mathbf{a_i}}$ represents the strong dissipation of the i-th cavity, and $\gamma \mathcal{L}_{\boldsymbol{\sigma}^x_i}$ is the weak bit flip process on the i-th qubit, which we have to counter. Towards achieving this, a particular coupling of the qubits to the cavities is achieved via the Hamiltonian, $\frac{\mathbf{H}(t)}{\bar{h}}$
\begin{IEEEeqnarray}{rCl}\label{eq:Hamiltonian}
	  =&  \sum\limits_{i=1} ^3 {\omega_{a_i}} \mathbf{a_i^{\dagger}a_i} + \sum\limits_{i=1} ^3 \frac{{\omega_{b_i}} }{2} {\boldsymbol{\sigma}^z_i} + \sum\limits_{i=1} ^3 \epsilon_i^a(t) (\mathbf{a_i^{\dagger}} + \mathbf{a_i}) \nonumber \\
	& + \sum\limits_{i=1} ^3 \epsilon_i^b(t) (\boldsymbol{\sigma}^+_i \text{+} \boldsymbol{\sigma}^-_i) - \sum\limits_{i=1} ^3 \frac{E_i}{\bar{h}} (\cos(\tfrac{\boldsymbol{\Phi}_i}{\Phi_0}) + \frac{1}{2} (\tfrac{\boldsymbol{\Phi}_i^2}{\Phi_0^2}) ) \; . 
\end{IEEEeqnarray}
Here the first two terms describe the dynamics of isolated cavities and qubits; the last one describes their nonlinear coupling as typically encountered in superconducting circuits with Josephson junctions; and the time dependent terms are drives on the cavities and the qubits (the pumps), which are of the form $\epsilon_1^a(t) = \epsilon_1^{a,1} (e^{i \omega_{p_1} t} + e^{-i \omega_{p_1} t}) + \epsilon_1^{a,2} (e^{i \omega_{p_2} t} + e^{-i \omega_{p_2} t})$. The reservoir will be engineered by selecting particular $\omega_{p_i}$,$\epsilon_1^{a,i}$, described later; the $\epsilon_i^b(t)$ are optional but would be of the same form, see below. Note that these are all fixed sinusoidal drives, without any precise feedback signals nor control logic to be timed.
  
The whole system is built to limit the effects of bit flips at rate $\gamma$: the cavities dissipating at rate $\kappa$ serve to evacuate the associated entropy and stabilize the system; the Hamiltonian construction must ensure the stabilization of the wanted subspace. The scheme differs from the standard one based on pairwise parity measurement. Indeed, each dissipative cavity is coupled to all the qubits and stabilizes the system as a whole.
  
 As ${\Phi_0^2} \gg 1$, we can simplify the last term by expanding the cosine to 4th order which gives
   \begin{IEEEeqnarray*}{rCl}
  \sum\limits_{i=1} ^3 \mathbf{a_i^{\dagger}a_i}( \frac{\chi_{a_i b_1}}{2}\boldsymbol{\sigma}^z_1 + \frac{\chi_{a_i b_2}}{2}\boldsymbol{\sigma}^z_2  + \frac{\chi_{a_i b_3}}{2}\boldsymbol{\sigma}^z_3) \\
+ \sum\limits_{i=1} ^3 K_{a_i a_i} {\mathbf{a}_i^\dagger}^2 \mathbf{a}_i^2 + \sum\limits_{i \ne j} ^3 K_{a_i a_j} \mathbf{a}_i^{\dagger} \mathbf{a}_i \mathbf{a}_j^{\dagger} \mathbf{a}_j +  \sum\limits_{i \ne j} ^3 K_{b_i b_j} {\boldsymbol{\sigma}^z_i\boldsymbol{\sigma}^z_j} \, .
\end{IEEEeqnarray*}

%\begin{dmath}\label{eq:Hamiltonian 2 }
%	\frac{\mathcal{H}(t)}{\bar{h}}  =  {\omega_{a_1}} \mathbf{a_1^{\dagger}a_1} + \sum\limits_{i=1} ^3 \frac{{\omega_{b_i}} }{2} {\boldsymbol{\sigma}^z_i} -  \mathbf{a_1^{\dagger}a_i}( \frac{\chi_{a_k b_1}}{2}\boldsymbol{\sigma}^z_1 + \frac{\chi_{a_k b_2}}{2}\boldsymbol{\sigma}^z_2  + \frac{\chi_{a_k b_3}}{2}\boldsymbol{\sigma}^z_3) + \epsilon_1^{a,1} (e^{i \omega_{p_1} t} + e^{-i \omega_{p_1} t}) + \epsilon_1^{a,2} (e^{i \omega_{p_2} t} + e^{-i \omega_{p_2} t})
%\end{dmath}

\subsubsection{Parameter tuning:} The reservoir is tuned by taking $\sum_{i=1}^3  \chi_{a_k b_i} = 0$ for all $k$, and $\omega_{p_1} = \frac{|\omega_{a_1} - \omega_{b_1}|}{2}$ , $ \omega_{p_2} = \frac{|\omega_{a_1} + \omega_{b_1}|}{2}$. The first condition ensures that the logical states ($|000\q$ and $|111\q$) of the qubit have the same energy. The second condition favors the conversion of a single qubit excitation into a decaying photon of the cavity, thanks to 2 pump photons at frequency $\omega_{p_1}$; and the re-excitation of a single decayed qubit simultaneously with the creation of a decaying photon in the cavity, by conversion of 2 pump photons at $\omega_{p_2}$. In both processes, the fast decay of the cavity photon inhibits the reverse process, which a pure Hamiltonian coupling would induce at equivalent rate.
Finally, we also fix the $\epsilon_1^{a,j}$ to satisfy 
\begin{equation}
\Omega_{p_j} :=\sqrt{K_{a_1 a_1} \chi_{a_1 b_1}} \mid \frac{\epsilon_1^{a,j}}{\omega_{a_1}-\omega_{p_j}} \mid ^2
= \Omega_p
\end{equation}
independently of $j$. The first and last condition are necessary for preserving any superposition of logical states $\alpha |000\q + \beta |111\q$.
In a first approach, we will keep the $\epsilon_i^b(t)=0$.

\subsubsection{Assumption:} The approach is based on turning couplings on or off by parametric resonance effects. For this, we assume the following timescale separation:
$\gamma \ll \kappa  \ll  \chi   \ll  \omega $. This is realistic in typical quantum superconducting circuits. Since cavity excitations are created through $\omega_{p_1}, \omega_{p_2}$ after bit-flip errors, having $n$ excitations in the cavities is proportional to $(\gamma/\kappa)^n$; this becomes negligible for $n>1$ in the regime $\gamma \ll \kappa$, and then the terms ${\mathbf{a}_i^\dagger}^2 \mathbf{a}_i^2$ and $\mathbf{a}_i^{\dagger} \mathbf{a}_i \mathbf{a}_j^{\dagger} \mathbf{a}_j$ vanish. The terms in ${\boldsymbol{\sigma}^z_i\boldsymbol{\sigma}^z_j}$ can be rigorously ignored by slightly modifying the pump frequencies.

\subsection{Error correction protocol}

An easy change of frame removes all the components of order $\omega$, the dominating rate. Then by performing a Rotating Wave Approximation (RWA) $\simeq$ averaging approximation on the basis of $\chi   \ll  \omega $, we obtain a simpler form of the Hamiltonian.
We then do a new change of rotational frame to remove the dominant terms now of order $\chi$, and a new RWA on the basis of $ \Omega_p  \ll  \chi$. Finally, choosing $\Omega_p < \kappa$, we use a last timescale separation to do adiabatic elimination: considering that the components in $\kappa$ quickly converge towards their stationary values, we eliminate the variables associated to the cavities and only study their effect on the slow dynamics, i.e.~the qubits. This gives the following effective master equation, describing how the engineered reservoir affects the three qubits composing the coding space:
  \begin{equation}\label{eq:reduitefull}
\frac{d \rho}{d t} =  \sum\limits_{i=1} ^3 \Gamma_c \mathcal{D}_{\mathbf{c_i}}(\rho) +  \gamma \mathcal{D}_{\boldsymbol{\sigma}^x_i}(\rho) \; .
\end{equation}
The second Lindblad terms formulate the bit flip errors; the first ones represent the effective error correction, induced by the operators $\mathbf{c}_1 = |000\q\p100| + |111\q \p011|$, $\mathbf{c}_2 = |000\q \p010| + |111\q \p010|$ and $\mathbf{c}_3 = |000\q \p001| + |111\q \p110|$. We can also write $\mathbf{c}_1 = \boldsymbol{\sigma}^-_1 \Pi_{|00\q}^{23} + \boldsymbol{\sigma}^+_1 \Pi_{|11\q}^{23} $ where $\Pi_{|00\q}^{23}$ is the projection operator of the second and third qubit on the
state $|00\q$. Moreover, $\Gamma_c$ represents the effective correction rate and is well approximated by $\Gamma_c = \frac{\Omega_p^2}{\kappa}$.
We thus achieve the protection of the logical qubit, as the natural effect associated to $\gamma$ is countered by the engineered effect associated to $\Gamma_c$.

\subsubsection{Implementation using only one cavity:} Instead of having one separate cavity to counter the bit-flip of each qubit, we can design an effective Hamiltonian which transfers the errors of the other qubits onto the first one. More precisely, we apply two extra drives $\epsilon_i^b(t)$ of fixed amplitudes and of frequencies $\omega_{p_{12}} = \frac{|\omega_{b_1} - \omega_{b_2}|}{2}$ and $\omega_{p_{23}} = \frac{|\omega_{b_2} - \omega_{b_3}|}{2}$.
These drives, together with the terms in ${\boldsymbol{\sigma}^z_i\boldsymbol{\sigma}^z_j}$, induce effective couplings which circulate the qubit states, such that correcting one ends up correcting all of them.

\section{Scaling up}

In discrete-time error correction based on projective parity measurement, the 3-qubit code allows to retrieve the correct information whenever one single qubit flips, but not when several ones flip together. To make a logical error less probable, an $n$-qubit code protects against up to $(n-1)/2$ qubits flipping simultaneously. A similar scaling is expected for information protection by reservoir engineering, and our purpose is to investigate how to implement it. The challenge is that in the above scheme, each cavity is coupled to all the qubits. Such one-to-all coupling is not realistically scalable from a physical engineering viewpoint.

We here propose a design allowing to scale up the number of qubits but without having to couple a cavity to an increasing number of qubits. For simplicity we describe the first level of scaling, from 3 towards 9 physical qubits.

\subsection{The star design}

\subsubsection{Description of the system:} 
\begin{figure}[ht]\centering
	\includegraphics[width=1\columnwidth]{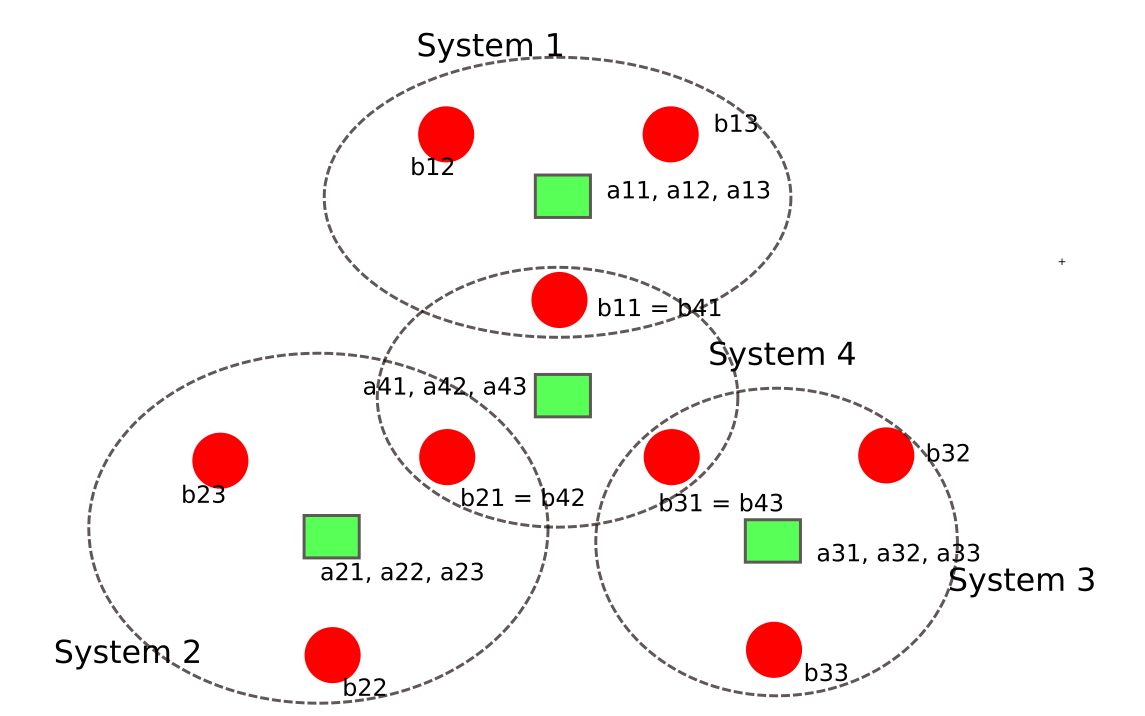}
	\caption{Star design for scaling up reservoir-engineering-based error correction. The qubits are represented as red discs; each black dotted circle delimits a ``subsystem'', i.e.~a set of qubits coupled to the cavity or cavities in its centre (green squares).}
	\label{schema}
\end{figure}

Our proposal consists of four instances of the system described in the last section, see Fig.~\ref{schema}. Each of the outer instances shares a common qubit with the central one. We thus have 9 qubits and 12 cavities (this number will be reduced later). We will note  $\mathbf{a_{ij}}$ the annihilation operator in the i-th system on j-th cavity, and similarly for the qubits.
The system is described by the Lindblad equation:
\begin{equation}\label{eq:Lindbladiannew}
\frac{d \rho}{d t} = -i[\mathbf{H}(t), \rho] + \sum\limits_{i=1} ^4\sum\limits_{j=1} ^3 \kappa \mathcal{D}_{\mathbf{a_{ij}}}(\rho) + \sum\limits_{i,j=1} ^3 \gamma \mathcal{D}_{\boldsymbol{\sigma}^x_{ij}}(\rho) 
\end{equation}
where the Hamiltonian, with the same simplifications and hypotheses as in the previous section, is described by $\frac{\mathbf{H}(t)}{\bar{h}} = \sum\limits_{i=1} ^4 \frac{\mathbf{H}_i(t)}{\bar{h}}$ with 
\begin{IEEEeqnarray*}{rCl}
\frac{\mathbf{H}_i(t)}{\bar{h}}	&=& \sum\limits_{j=2} ^3 \epsilon_{ij}^a(t)  (\mathbf{a_{ij}}  + h.c.) + \sum\limits_{j=1} ^3  \Big[ {\omega_{a_{ij}}} \mathbf{a_{ij}^{\dagger}a_{ij}}  + \frac{{\omega_{b_{ij}}} }{2} \\
	 && - \mathbf{a_{ij}^{\dagger}a_{ij}}( \frac{\chi_{a_{ij} b_{i1}}}{2}\boldsymbol{\sigma}^z_{i1} + \frac{\chi_{a_{ij} b_{i2}}}{2}\boldsymbol{\sigma}^z_{i2}  + \frac{\chi_{a_{ij} b_{i3}}}{2}\boldsymbol{\sigma}^z_{i3})  \nonumber \Big] +  \\
	&& +  \epsilon_{i1}^a(t) \Big(  \mathbf{a_{i1}} (1 + \sum\limits_{k=1} ^3 e^{i \chi_{a_{4k} b_{i1} t}} +  e^{-i \chi_{a_{4k} b_{i1} t}} ) + h.c. \Big)
\end{IEEEeqnarray*}
%\begin{IEEEeqnarray*}{Cl}
%	  =&  \sum\limits_{j=1} ^3  \Big[ - \mathbf{a_{ij}^{\dagger}a_{ij}}( \frac{\chi_{a_{ij} b_{i1}}}{2}\boldsymbol{\sigma}^z_{i1} + \frac{\chi_{a_{ij} b_{i2}}}{2}\boldsymbol{\sigma}^z_{i2}  + \frac{\chi_{a_{ij} b_{i3}}}{2}\boldsymbol{\sigma}^z_{i3}) \\
%%
%& \quad +\>  {\omega_{a_{ij}}} \mathbf{a_{ij}^{\dagger}a_{ij}}  + \frac{{\omega_{b_{ij}}} }{2}   \nonumber \Big] + \sum\limits_{j=2} ^3 \epsilon_{ij}^a(t)  (\mathbf{a_{ij}}  + h.c.) \\
%& +  \epsilon_{i1}^a(t) \Big(  \mathbf{a_{i1}} (1 + \sum\limits_{k=1} ^3 e^{i \chi_{a_{4k} b_{i1} t}} +  e^{-i \chi_{a_{4k} b_{i1} t}} ) + h.c. \Big)
%\end{IEEEeqnarray*}
for $i = 1,2,3$, and $\frac{\mathbf{H}_4(t)}{\bar{h}} $
\begin{IEEEeqnarray*}{rCl}
	=&  \sum\limits_{j=1} ^3  \Big[ \mathbf{a_{4j}^{\dagger}a_{4j}}( \frac{\chi_{a_{4j} b_{11}}}{2}\boldsymbol{\sigma}^z_{11} + \frac{\chi_{a_{4j} b_{21}}}{2}\boldsymbol{\sigma}^z_{21}  + \frac{\chi_{a_{4j} b_{31}}}{2}\boldsymbol{\sigma}^z_{31}) \\
	& \quad -\>  {\omega_{a_{4j}}} \mathbf{a_{4j}^{\dagger}a_{4j}}  + \frac{{\omega_{b_{j1}}} }{2}   \nonumber \Big] \\
	& + \sum\limits_{j=1} ^3 \epsilon_{4j}^a(t)  \Big(  \mathbf{a_{4j}} (1 + \sum\limits_{k=1} ^3 e^{i \chi_{a_{jk} b_{j1} t}} +  e^{-i \chi_{a_{jk} b_{j1} t}} ) + h.c. \Big) 
\end{IEEEeqnarray*}
The new terms, involving drives of the form,
\begin{equation}\label{new}
\epsilon_{i1}^a(t)\sum\limits_{k=1} ^3 e^{i \chi_{a_{4k} b_{i1} t}} +  e^{-i \chi_{a_{4k} b_{i1} t}}  \; , 
\end{equation}
are added to deal with the effects of overlapping subsystems. For example, if the cavity $a_{11}$ needs to correct the qubit $b_{11}$, it needs to do it both when the cavities $a_{41}$, $a_{42}$ and $a_{43}$ are populated or not. These possibilities lead to more energy levels for which we want the correction to take place, and thus more transition frequencies to be activated. Choosing two coupling strengths equal in each subsystem, for example $\chi_{a_{ij} b_{k1}} = \chi_{a_{ij} b_{k2}}  = -\frac{\chi_{a_{ij} b_{k3}}}{2}$, allows to only add 2 extra drives instead of 3.

The parameters in each subsystem are tuned similarly to the building block presented in the previous section. The only new requirement is that the central subsystem must have different coupling terms ($\chi$) from the outer ones, in order to prevent some unwanted transitions.

\subsubsection{Error correction protocol:} Similarly to section 2, we perform a first change of frame to remove all the components of order $\omega$, allowing to perform the standard RWA in the $\chi  \ll  \omega $ regime; then a second change of frame, a new RWA and a model reduction, to obtain the final equation approximating the dynamics for large time-scale separation:
\begin{equation}\label{eq:reduitefullstar}
\frac{d \rho}{d t} =  \sum\limits_{i=1} ^4\sum\limits_{j=1} ^3 \Gamma_c^i \mathcal{D}_{\mathbf{c_{ij}}}(\rho) +  \sum\limits_{i=1} ^3\sum\limits_{j=1} ^3\gamma \mathcal{D}_{\boldsymbol{\sigma}^x_{ij}}(\rho) \; . 
\end{equation}
Here $\mathbf{c_{ij}} = (\boldsymbol{\sigma}^-_{ij} \Pi_{i_{|00\q}}^{ \neq j} + \boldsymbol{\sigma}^+_{ij} \Pi_{i_{11\q}}^{ \neq j}) \otimes \boldsymbol{I_{ \neq i}}$    for $i = 1,2,3$, where $\Pi_{i_{|00\q}}^{ \neq j}$ is the projection on $|00\q$ for the two qubits different from $j$ in the $i$-th subsystem. $\boldsymbol{I_{ \neq i}}$ is the identity operator on the qubits not belonging to the $i$-th system. We use a convenient notation where qubit 1 of subsystem $i \in \{1,2,3\}$ is the same entity as qubit $i$ of subsystem 4.

For $i \in \{1,2,3\}$, the Lindblad equation \eqref{eq:reduitefullstar} expresses the correction of single bit-flips, just like in the building block of Section 2. In addition, subsystem 4 performs a similar correction towards the span of $|000\rangle,|111\rangle$ among qubits of the $i \in \{1,2,3\}$ subsystems. Together, this enforces all qubits to agree.

%(maybe give example $\mathbf{c_{11}} = (\boldsymbol{\sigma}^-_{11} \Pi_{1_{|00\q}}^{23} + \boldsymbol{\sigma}^+_{11} \Pi_{1_{|00\q}}^{23}) \otimes \boldsymbol{I_{2,3}}$)

%We have also $\mathbf{c_{4j}} = (\boldsymbol{\sigma}^-_{j1} \Pi_{i_{|00\q}}^{k \neq j} + \boldsymbol{\sigma}^+_{ij} \Pi_{i_{|00\q}}^{k \neq j}) \otimes \boldsymbol{I_{l \neq i}}$

\subsubsection{Implementation using less cavities:} Similarly as in section 2, we could implement a single cavity in each subsystem and add drives to circulate the qubit states.

\subsubsection{Alternative:} Instead of adding the new drives (\ref{new}), an alternative solution would be to periodically turn on the outer subsystems correction while the inner one is off, and reciprocally. This totally cuts off the need of adding new drives, as at each time instant the active correction protocol is not perturbed by any overlapping subsystem. This leaves more room to have separate frequencies towards validating the RWA. While this requires an additional external intervention, it does not require any precise timing unlike feedback based error correction.

\subsubsection{Convergence result:} For $\gamma=0$, i.e.~in absence of perturbation, these schemes based on \eqref{eq:reduitefullstar} exponentially stabilize the logical code space $\text{span}(|0\q^{\otimes 9} ,\; |1\q^{\otimes 9} )$, at a rate proportional to $\min_i \, \Gamma_c^i$. Furthermore, any perturbed initial state $c_0 \, |\psi_0\q + c_1 \, |\psi_1\q$ where $\psi_0$ is a linear combination of basis states with at most 3 qubits on $|1\q$, and $\psi_1$ is a linear combination of basis states with at most 3 qubits on $|0\q$, gets mapped to $c_0 |0\q^{\otimes 9} + c_1 |1\q^{\otimes 9}$.

For errors appearing continuously, i.e.~$\gamma > 0$, the induced perturbation on logical states should be small, thanks to robustness of exponential stabilization. The precise benefits of the various alternatives is under investigation. Like for measurement-based error correction, a threshold value for $\gamma$ should be found, such that when $\gamma$ is very small it is beneficial to scale up the scheme, while when $\gamma$ is too high we cannot improve information protection by adding more qubits. For measurement-based error correction, such thresholds are usually found by extensive simulation, except in idealized cases. We hope that the continuous-time setting and systems theory tools could lead to (approximate) analytic bounds.

\section{Simulation results}

To get first insights, we show different simulations of the master equation from section 2, after performing only the first RWA. We plot the fidelity of the state over time to the initial state,
\begin{equation}\label{eq:fid}
F(t) = \text{trace}(\rho(0) \rho (t))
\end{equation}
for $\rho(0)=|\psi_0\q\p\psi_0|$ with $|\psi_0\q = \frac{|000\q - |111\q}{\sqrt{2}}$.

\begin{figure}[ht]\centering 
	\includegraphics[width=0.8\columnwidth]{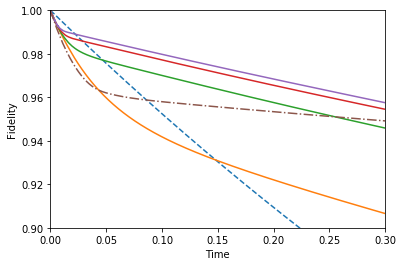}
	\caption{Orange line: $\Omega_p = 100\gamma$. Green line: $\Omega_p = 200\gamma$. Red line: $\Omega_p = 300\gamma$. Violet line: $\Omega_p = 400\gamma$. Dashed blue line: single qubit without protection. Brown dashdotted line: single cavity protocol.}
	\label{toto}
\end{figure}

For all the simulations we have taken $\gamma = 1, \kappa = 500\gamma$. The dispersive couplings are chosen to satisfy $\chi_{a_k b_2} = \chi_{a_k b_3} = -\frac{\chi_{a_k b_1}}{2}$ for the three cavities. We take $\chi_{a_k b_1}= 100 \Omega_p$ towards satisfying the second RWA; this is confirmed in the simulations as they work as expected. The last model reduction, i.e.~adiabatic elimination, would require $\Omega_p < \kappa$ in order to ensure information protection at a rate $\Omega_p^2 / \kappa$; this expresses the physical fact that the cavity cannot evacuate entropy at a rate faster than $\kappa$. This effect is visible on Fig. \ref{toto}, as we observe that an augmentation of $\Omega_p$ results in a higher fidelity over time, with a saturation when it reaches $O(\kappa)$. The fidelity resulting from this scheme is above the one of an unprotected physical qubit, confirming efficient error protection. Figure \ref{toto} also compares the 3-cavity protocol with the one using a single cavity and qubit circulation drives.
 At initial times, as expected, the 3 cavities evacuate entropy faster than the single one and provide better protection. However, at longer times, the single cavity protocol loses information more slowly. This is due to the fact that it better satisfies the second RWA.

The 9-qubit star design, because of its high dimension ($2^{9+4}$ even for the simplified implementation), has not been fully simulated yet but will be available soon. We expect that the slope of fidelity loss will be even flatter, and preliminary results are indeed going in this direction.

\section{Conclusion}
To sum up, we have taken a reservoir engineering method based on the coupling of 3 qubits to cavities, and have shown how to use this system as a building block in a star design using more physical qubits to protect more strongly our logical qubit, while avoiding all-to-all coupling. We have done so by using the Rotative Wave Approximation and adiabatic elimination, taking advantage of the several orders of timescales in our physical system.

It is possible to continue scaling up our system by adding building blocks at the extremities of all the outer qubits. Future work will try to find analytical boundaries on the error correction rate and how it would scale with the number of building blocks.

\end{document}